\documentclass[twocolumn,showpacs,preprintnumbers,superscriptaddress,amsmath,amssymb]{revtex4}
\usepackage{graphicx}% Include figure files
\usepackage{dcolumn}% Align table columns on decimal point
\usepackage{bm}% bold math
\usepackage{color}

\begin{document}

\title{A New Type of Plasma Wakefield Accelerator Driven by Magnetowaves}
\author{Pisin Chen}%
\email{chen@slac.stanford.edu} \affiliation{Kavli Institute for
Particle Astrophysics and Cosmology, Stanford Linear Accelerator
Center, Stanford University, Stanford, CA 94305,
USA.}\affiliation{Department of Physics and Graduate Institute of Astrophysics, National Taiwan
University, Taipei 106, Taiwan.}\affiliation{Leung Center for Cosmology and Particle Astrophysics, National Taiwan University, Taipei 106, Taiwan.}
\author{Feng-Yin Chang}
\affiliation{Institute of
Physics, National Chiao-Tung University, Hsinchu 300, Taiwan.}\affiliation{Leung Center for Cosmology and Particle Astrophysics, National Taiwan University, Taipei 106, Taiwan.}
\author{Guey-Lin Lin}
\affiliation{Institute of Physics, National Chiao-Tung University,
Hsinchu 300, Taiwan.}\affiliation{Leung Center for Cosmology and Particle Astrophysics, National Taiwan University, Taipei 106, Taiwan.}
\author{Robert J. Noble}
\affiliation{Stanford Linear Accelerator Center, Stanford
University, Stanford, CA 94305, USA.}
\author{Richard Sydora}
\affiliation{Department of Physics, University of Alberta,
Edmonton, Alberta, Canada.}

%\date{\today}
\begin{abstract}
We present a new concept for a plasma wakefield accelerator driven by magnetowaves (MPWA). This concept was originally proposed as a viable mechanism for the "cosmic accelerator" that would accelerate cosmic particles to ultra high energies in the astrophysical setting.  Unlike the more familiar Plasma Wakefield Accelerator (PWFA) and the Laser Wakefield Accelerator (LWFA) where the drivers, the charged-particle beam and the laser, are independently existing entities, MPWA invokes the high-frequency and high-speed whistler
mode as the driver, which is a medium wave that cannot exist outside of the plasma. Aside from the difference in drivers, the underlying mechanism that excites the plasma wakefield via the ponderomotive potential is common. Our computer simulations show that under
appropriate conditions, the plasma wakefield maintains very high
coherence and can sustain high-gradient acceleration over many plasma wavelengths. We suggest that in addition to its celestial application, the MPWA concept can also be of terrestrial utility. A proof-of-principle experiment on MPWA would benefit both terrestrial and celestial accelerator concepts.
\end{abstract}
\maketitle
Ultra high energy cosmic rays (UHECR) with energy beyond $10^{20}$eV have been observed, yet its astrophysical  source and acceleration mechanism remain a mystery. This has been considered as one of the 11 most important questions in cosmology and particle astrophysics in the 21st century\cite{NRC}. Plasma wakefield accelerators\cite{LWFA:Tajima,PWFA:Chen85,plasma:Esarey} are
known to possess two salient features: (1) The energy gain per unit
distance does not depend on the particle's instantaneous
energy or momentum. This is essential to avoid the gradual decrease
of efficiency in reaching ultra high energies; (2) The acceleration
is linear. Bending of the trajectory is not a necessary condition for this
mechanism. This helps to minimize inherent energy loss which would
be severe at ultra high energy. These qualities suggest that plasma wakefield might be a good candidate for the cosmic accelerator. However, high-intensity, ultra-short
photon or particle beam pulses that excite the laboratory plasma
wakefields are not readily available in the astrophysical setting.
In 2002 it was proposed\cite{MPWA:Chen02} that large amplitude
plasma wakefields can instead be excited by the astrophysically more
abundant plasma ``magnetowaves", whose field components are magnetic
in nature ($|B|>|E|$). As astrophysical outflows are most likely random, protons may be stochastically accelerated beyond ZeV energy by riding on such random wakefields, which would explain the observed UHECR inverse-power-law energy spectrum. Recent particle-in-cell simulations have demonstrated that this MPWA mechanism is indeed valid\cite{MPWA:Chang08}.

In this paper we suggest that in addition to its application to the celestial accelerator, this new way of plasma wakefield excitation may also be applicable to high acceleration-gradient terrestrial accelerators. In the case of the electron-beam-driven Plasma Wakefield Accelerator (PWFA)\cite{PWFA:Chen85} and the laser-driven Laser Wakefield Accelerator (LWFA)\cite{LWFA:Tajima}, the drivers are physical entities that can independently exist outside of plasma. In contrast the driver of the MPWA, the magnetowave, is a ``medium wave" which cannot exist without the plasma medium. The production of ultra short and high intensity laser and electron beams requires complex and expensive facilities. Magnetowaves on the other hand can be excited easily by transversely perturbing a background static magnetic field. This salient feature would hopefully be an attractive alternative to drive the plasma wakefield accelerator for laboratory applications. A proof-of-principle experiment on MPWA would not only benefit the terrestrial accelerator technology but also contribute to the understanding of celestial acceleration.

Magnetized plasmas support a variety of wave modes propagating at
arbitrary angles to the imposed magnetic field. For our purpose, we
focus on wave propagation parallel to the external magnetic
field to ensure the linear acceleration. In this case, the
electromagnetic waves become circularly polarized and the dispersion
relation in the non-relativistic limit is \cite{Stix}
\begin{equation}
 \omega^{2}=k^{2}c^{2}+\frac{\omega_{ip}^{2}}{1\pm\omega_{ic}/\omega}+
\frac{\omega_{ep}^{2}}{1\mp\omega_{ec}/\omega}\, ,\label{dispersion}
\end{equation}
where the upper (lower) signs denote the right-hand (left-hand)
circularly polarized waves. $\omega_{ip,ep}=\sqrt{4\pi e^2 n_{p}/m_{i,e}}$
is the ion and electron plasma frequency and  $\omega_{ic,ec} = eB/m_{i,e}c$ the ion and electron cyclotron frequency, respectively. Each polarization has two real solutions with high and
low frequency branches and both have a frequency cutoff which
forms a forbidden gap for wave propagation. The right-hand
polarized, low frequency solution is called the whistler wave
which propagates at a phase velocity less than the speed of light.
When the magnetic field is sufficiently strong such that
$\omega_{ec}\gg \omega_{ep}$, the dispersion of the whistler mode
becomes more linear over a wider range of wavenumbers with phase
velocity approaching the speed of light (see Fig.1). The $E$ and
$B$ components of the wave are now comparable in strength. In this
regime the traveling wave pulses can maintain their shape over
macroscopic distance, a condition desirable for plasma wakefield
acceleration.
\begin{figure}[tb]
\begin{center}
$\begin{array}{c}
\includegraphics[width=8.5cm]{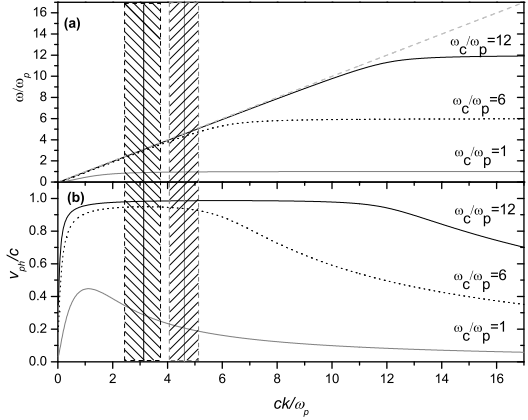}
\end{array}$
\end{center}
\caption{(a) Frequency and (b) phase velocity versus wavenumber
for different magnetic field strengths. The vertical solid lines are
the mean values of the pulse wavenumbers that were chosen for the PIC
simulation for Case 1 and Case 2, and the shaded regions their ranges.} \label{pulse}
\end{figure}

Applying the dispersion relation for the whistler wave, with the ion
motion neglected, we obtain the ponderomotive force acting on an
individual electron as\cite{wk}
\begin{equation}
F_{z}=-\frac{1}{2}\frac{e^2}{m_e\omega
(\omega-\omega_{c})}\left[1+\frac{kv_g\omega_c}{\omega(\omega-\omega_c)}\right]\partial_{\zeta}E_{_W}^2(\zeta),
\end{equation}
where $E_{_W}(\zeta)$ is the amplitude of the whistler wave-packet,
and $\zeta\equiv z-v_gt$ the co-moving coordinate for the driving
pulse. From here on, we drop all the indices for ion and electron. Combining this equation with the continuity equation and the
Poisson equation, the longitudinal electric field in the plasma,
i.e., the plasma wakefield, can be solved and it reads
\begin{eqnarray}
E_z(\zeta)=-\frac{ek_pE_{_W}^2}
{m_e\omega(\omega-\omega_c)}\left[1+\frac{kv_g\omega_c}{\omega(\omega-\omega_c)}\right]
\chi(\zeta), \label{wakefield}
\end{eqnarray}
with
\begin{equation}
\chi(\zeta)=\frac{k_p}{2E_{_W}^2}\int_{\zeta}^{\infty}d\zeta^{\prime}E_{_W}^2(\zeta^{\prime})
\cos\left[k_p(\zeta-\zeta^{\prime})\right],
\end{equation}
where $E_{_W}$ is the maximum value of $E_{_W}(\zeta)$. An
expression similar to Eq.~(\ref{wakefield}) has been obtained for
laser-induced wakefield in a magnetized plasma\cite{sh}. For a
Gaussian driving pulse with $E_{_W}(\zeta)=E_{_W}
\exp(-\zeta^2/2\sigma^2)$, it can be shown that behind the driving
pulse, i. e., $|\zeta| \gg \sigma$,
\begin{eqnarray}
\chi(\zeta)=\frac{\sqrt{\pi}}{2} k_p\sigma
e^{-k_p^2\sigma^2/4}\cos k_p\zeta \equiv \chi\cos k_p\zeta.
\end{eqnarray}
It is customary to express the plasma wakefield in terms of the
Lorentz invariant ``strength parameter" of the driving pulse,
$a_0\equiv eE_{_W}/m_ec\omega$, and the ``cold wavebreaking" field,
$E_{wb}\equiv m_ec\omega_p/e$. Assuming the driving pulse frequency
is centered around $\omega$ and its speed $v_{g}\approx \omega/k
\sim c$, the maximum wakefield, or the {\it acceleration gradient},
attainable behind the driving pulse is then
\begin{equation}
G=
\frac{k^{2}c^2}{\omega^2}\frac{a_0^2}{(1-\omega_c/\omega)^2}\chi eE_{wb}\approx \frac{a_0^2}{(1-\omega_c/\omega)^2}\chi eE_{wb}.
\label{wakefield4}
\end{equation}
We see that in the non-relativistic limit, i.e., the linear regime of plasma perturbation, the wakefield in a magnetized plasma is enhanced over the unmagnetized case by a factor $1/(1-\omega_c/\omega)^2$, which can be substantial when $\omega$ approaches $\omega_c$.

We have conducted computer simulations to study the MPWA process
driven by a Gaussian whistler pulse described above. Our simulation
model integrates the relativistic Newton-Lorentz equations of motion
in the self-consistent electric and magnetic fields determined by
the solution to Maxwell's equations\cite{Dawson,Sydora}. The
4-dimensional phase space $(z,p_x,p_y,p_z)$ is used for the charged
particle dynamics and a uniform external magnetic field, $B_0$, is
imposed in the z-direction. In order to investigate the influence of magnetization on the wakefield excitation, we study two cases with the normalized
physical parameters $\omega_{c}/ \omega_{p}=6$ (Case 1) and 12 (Case 2).
We used a wavepacket with Gaussian width
$\sigma =80\Delta/\sqrt{2}$, where $\Delta$ is the cell size taken
to be unity, and we assigned the wavenumber $k = 2\pi/60\Delta$ and $ 2\pi/40\Delta$ to Case 1 and Case 2, respectively.
All other basic parameters are common, i.e., we set $m_i/m_e=2000$ and imposed a uniform background plasma with electron collisionless skin depth, $c/\omega_{p}=30\Delta$. This gives
$\omega/\omega_{p}=2.98$ and $v_{g}/c\simeq \omega/ck= 0.95$ for Case 1 and $\omega/\omega_{p}=4.64$ and $v_{g}/c\simeq \omega/ck= 0.99$ for Case 2. Other
numerical parameters used were: total number of cells in the
$z$-direction, $L_z =2^{14}\Delta\simeq 546c/\omega_p$, average number of
particles per cell was 10, and the time step $\omega_p\Delta t=0.1$ for Case 1 and 0.05 for Case 2.
The fields were normalized by $(1/30)E_{wb}$.

To compare the two cases on the equal footing, we choose to fix the gradient $G$. This is done by setting the maximum amplitude for Case 1 at $E_{_W}=8.15$ and, commensurate with the increase in $\omega$, the maximum amplitude for Case 2 at $E_{_W}=20$. These  give the normalized
vector potential $a_{0}=eE_{_W}/m_ec\omega=0.09$ for Case 1 and 0.14 for Case 2. Thus the
wakefield in our simulation is in the linear regime. The pulse was
initialized at $z_0=500\Delta=16.66 c/\omega_p$. To avoid spurious
effects, we gradually ramped up the driving pulse amplitude until
$t=100 \omega_{p}^{-1}$ for Case 1 and $t=200 \omega_{p}^{-1}$ for Case 2, during which the plasma feedback to the
driving pulse was ignored. After these times, the driving pulse-plasma
interaction was tracked self-consistently. As the dispersion
relation in this regime is not perfectly linear, there was a gradual
spread of the pulse width. Thus $\chi$ and $E_{_W}$ of the driving
pulse decreased accordingly. As a result, the maximum wakefield
amplitude, $E_z$, declined in time. We shall see, however, that such degradation becomes much milder as the ratio of $\omega_{c}/ \omega_{p}$ increases. Fig.2
is a snapshot of $E_{x}$ and $E_{z}$ in Case 1 at $\Delta t=100\omega_{p}^{-1}$ after the pulse was released. The initial wakefield amplitude agrees well with the theoretical maximum of $E_{z}\sim 0.0059 E_{wb}$. Fig.3 shows the snapshot of the same run at the later time $\Delta t=300\omega_{p}^{-1}$. Here we see the obvious dispersion of the driving pulse. We note, however, that while the driving pulse continues to disperse, the
wakefield remains coherent albeit with a much lower amplitude.
\begin{figure}[htb]
\begin{center}
$\begin{array}{c}
\includegraphics[width=8.5cm]{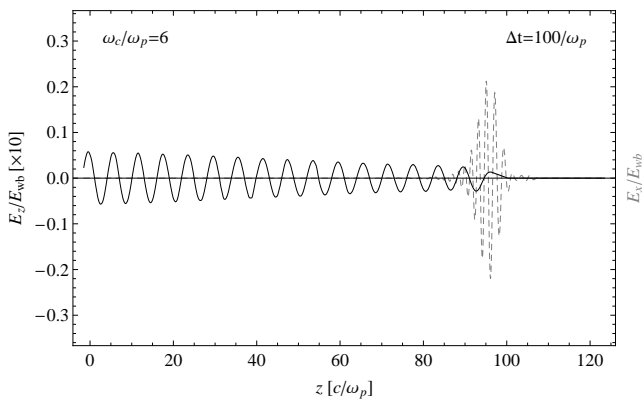}
\end{array}$
\end{center}
\caption{A snapshot of the plasma wakefield, $E_z$ (in black), induced
by the whistler pulse, $E_x$ (in gray) for Case 1 at the time step $\Delta t=100\omega_{p}^{-1}$.}
\label{wakeplot6a}
\end{figure}
\begin{figure}[tb]
\begin{center}
$\begin{array}{c}
\includegraphics[width=8.5cm]{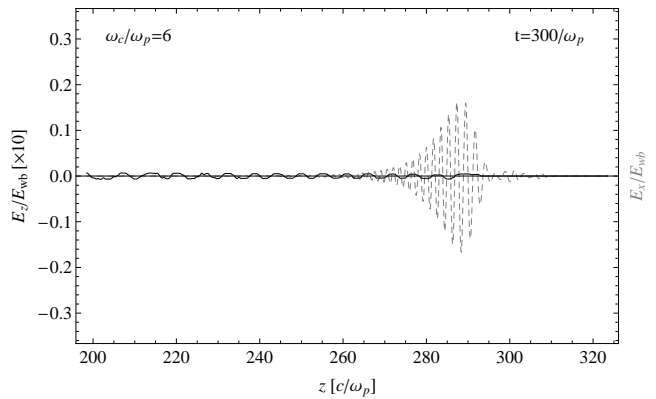}
\end{array}$
\end{center}
\caption{A late time snapshot for Case 1 at $\Delta t=300\omega_{p}^{-1}$.}
\label{wakeplot6b}
\end{figure}

Next we repeat the exercise with Case 2. Fig.4 shows the snapshot of Case 2 at  $\Delta t=100\omega_{p}^{-1}$. Comparing with Case 1 at the same time step, we see that the wakefield amplitudes in Case 2 had barely degraded and were more constant than that in Case 1, which is indicative that the driving pulse had smaller dispersion for Case 2. Fig.5 shows the snapshot of Case 2 at  $\Delta t=300\omega_{p}^{-1}$. Here the difference is more dramatic. Whereas the driving pulse had already widely spread and the corresponding wakefield had largely diminished in Case 1 by this time step, that in Case 2 were still well maintained.
\begin{figure}[tb]
\begin{center}
$\begin{array}{c}
\includegraphics[width=8.5cm]{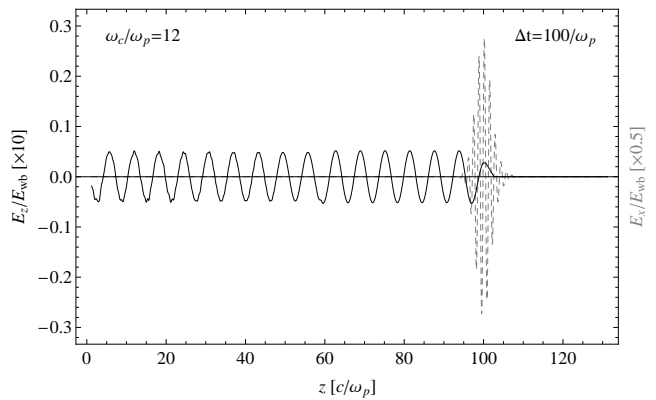}
\end{array}$
\end{center}
\caption{A snapshot of the plasma wakefield, $E_z$ (in black), induced
by the whistler pulse, $E_x$ (in gray) for Case 2 at the time step $\Delta t=100\omega_{p}^{-1}$.}
\label{wakeplot12a}
\end{figure}
\begin{figure}[tb]
\begin{center}
$\begin{array}{c}
\includegraphics[width=8.5cm]{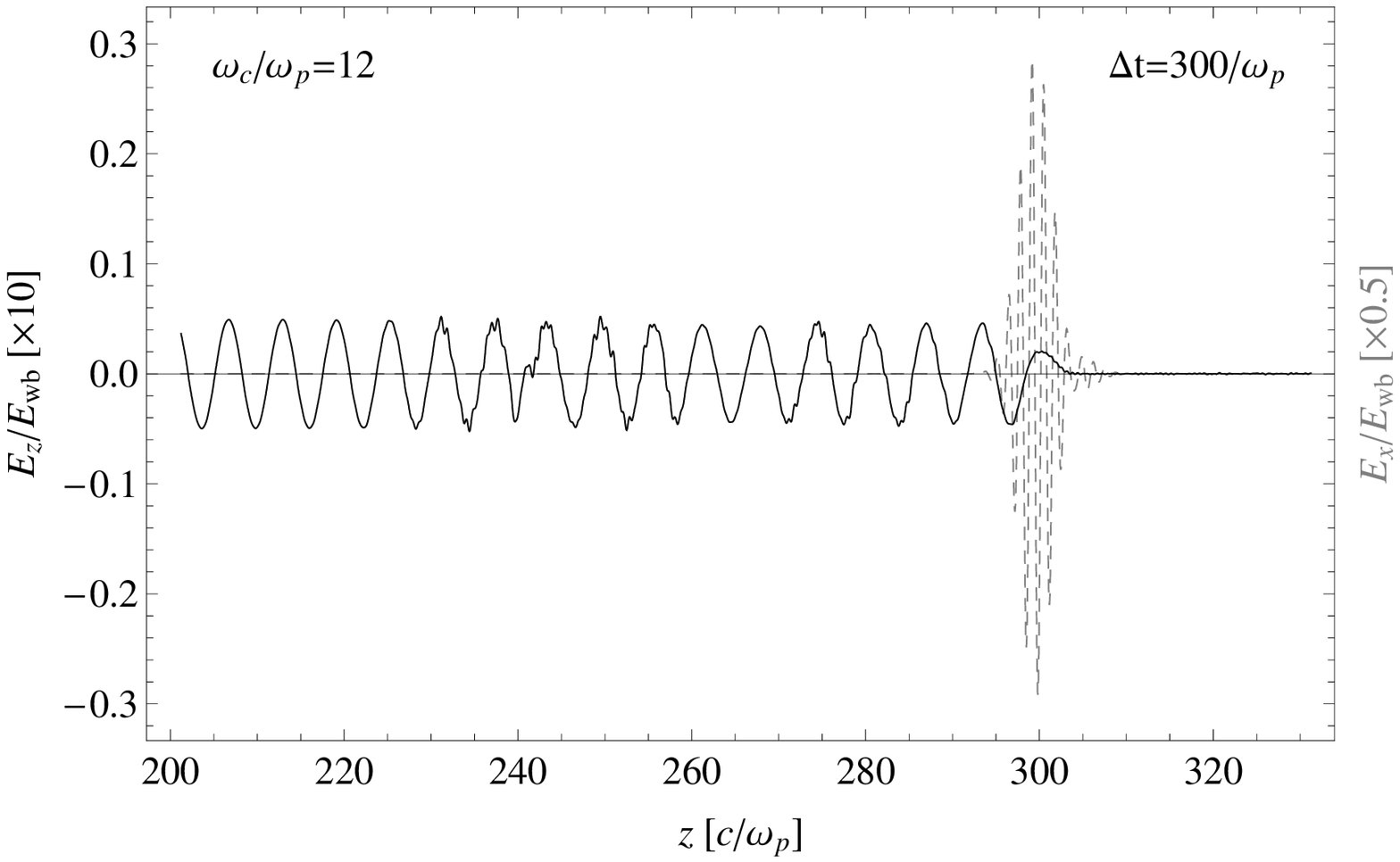}
\end{array}$
\end{center}
\caption{A late time snapshot for Case 2 at $\Delta t=300\omega_{p}^{-1}$.}
\label{wakeplot12b}
\end{figure}

\begin{figure}[tb]
\begin{center}
$\begin{array}{c}
\includegraphics[width=8.5cm]{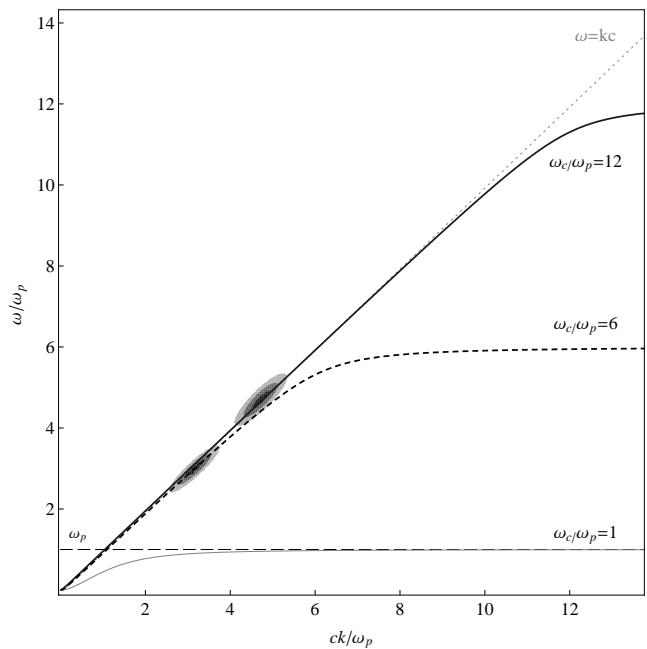}
\end{array}$
\end{center}
\caption{The intensity contours of the driving pulse as a function of ($\omega,k$)
from PIC simulation. The light cone and the linear dispersion curves for the whistler wave with $\omega_{c}/ \omega_{p}=1$, 6 and 12 are superimposed.}\label{simdisp}
\end{figure}

In Fig.\ref{simdisp}, we show the PIC simulations driving pulse intensity contours in the $\omega -k$ space for the two cases. There are superimposed with the theoretical curves for the whistler wave dispersion relations deduced from Eq.(1). We confirm that our driving pulses were indeed whistler waves.

In our previous paper\cite{MPWA:Chang08}, we have confirmed via PIC simulations the concept of plasma wakefield
excited by a magnetowave in a magnetized plasma. In this paper we further demonstrate that by raising the $\omega_{c}/ \omega_{p}$ ratio, the driving whistler pulse becomes less vulnerable to the dispersion. This aspect is especially important for MPWA to be a viable mechanism for terrestrial accelerator since it is essential for an accelerated particle to continuously gain energy from the plasma wakefield so as to attain high energy. Different from
the laser and the particle beam, the magnetowave is a medium wave
which cannot exist without the support from the plasma. MPWA should thus be of
interest as a fundamental phenomena in plasma physics and an
alternative approach to plasma wakefield acceleration.

Aside from the issues of dispersion of the driving pulse and sustainment of the plasma wakefield, we have not yet investigated other key accelerator physics issues in MPWA. It is curious whether MPWA would be more or less sensitive to plasma uniformity than PWFA and LWFA do. With regard to the phase slippage between the driving pulse and the accelerated particle, we see from our simulation results that by raising $\omega_{c}/ \omega_{p}$ from 6 to 12, the phase velocity of the wakefield increases from 0.95 to 0.99. This indicates that phase velocities can approach the speed of light if this ratio is further raised. One challenge common to plasma wakefield accelerators is that, in order to concentrate the driving pulse energy, its transverse size is often as small as a few plasma skin-depths. This is known to induce large transverse wakefield which in turn creates problems with undesirably large betatron motions for the accelerated particle. It would be interesting to know if this new type of plasma wakefield driver using the magnetowaves can help to ameliorate this difficulty.

Finally, it would be extremely exciting if proof-of-principle experiments on MPWA can be pursued. With regard to the possible physical mechanism to excite the whistler magnetowave driving pulse for experimentation, it has been shown that a fast ion-acoustic wave can decay into a whistler wave plus an ion-acoustic wave\cite{Bharuthram:1988}. It is therefore conceivable that such a decay process, or conversely the fusion of two ion-acoustic waves, can produce whistler wave. Inspired by this, one wonders whether a similar process can occur between a light wave and a whistler wave. If so, then perhaps a laser pulses may convert into a whistler wave pulse in a magnetized plasma under suitable conditions.

This work is supported by US DOE (Contract No.
DE-AC03-76SF00515), National Science Council of Taiwan (Grant No.
95-2119-M-009-026), and Natural Sciences and Engineering Research
Council of Canada.

\end{document}